\documentstyle[aps,epsfig,floats]{revtex}

\def\DESepsf(#1 width #2){\epsfxsize=#2 \epsfbox{#1}}
\def \beq{\begin{equation}}
\def \eeq{\end{equation}}
\def \be{\begin{eqnarray}}
\def \ee{\end{eqnarray}}
\def \bea{\begin{array}}
\def \eea{\end{array}}

\begin{document}
\draft


%
\twocolumn[\hsize\textwidth\columnwidth\hsize\csname @twocolumnfalse\endcsname

\title{$T$ violation in $B\rightarrow K^* \ell^+
\ell^-$ from SUSY}
\author{{ Chuan-Hung Chen$^{a,b}$ and C.~Q.~Geng$^{c,d}$
}
\\
{\small $^a$Department of Physics, National Cheng Kung
University
  Tainan, Taiwan,  Republic of China \\
  $^b$Institute of Physics, Academia Sinica,
  Taipei, Taiwan 115,  Republic of China \\
$^c$Department of Physics, National Tsing Hua University,
Hsinchu, Taiwan, Republic of China\\
$^d$Theory Group, TRIUMF,
4004 Wesbrrok Mall, Vancouver, B.C. V6T 2A3, Canada }}

\maketitle
\draft

\begin{abstract}
We propose to use T-odd momentum correlations as physical
observables to study T violation in $B\to K^{*}l^{+}l^{-}\ (l=e,\mu)$
decays. We show that these observables are zero in the standard model
but sizable at the level of $10\%$ in the supersymmetric
model. These large T violating effects are
measurable at the B-factories in KEK and SLAC and
future hadron colliders.
We also point out that the T violating effects are insensitive
to nonperturbative QCD contributions.
\end{abstract}
 ]


The study of flavor change neutral current (FCNC) in B decays has an
enormous progress since the CLEO observation \cite{cleo} of the radiative
$b\to s\gamma $ decay.
 Recently, the process of $B\to K\mu^+\mu^-$ has been
observed \cite{Belle} at the Belle detector in the KEKB $e^+e^-$ storage
ring with the branching ratio of
 $(7.5^{+2.5}_{-2.1}\pm0.9)\times 10^{-7}$, while
the SM expectation is around $5\times 10^{-7}$ \cite{Bdecays}.
Experimental searches at the
B-factories for $B\to K^*l^+l^-$ are also
within the theoretical predicted ranges \cite{BCP43}.

Via $B$ decays such as $B\to J/\Psi\,K$, we can
test the origin of CP violation (CPV) in the SM
which is the consequence of the CKM quark mixing matrix \cite{CKM}.
CP asymmetries (CPAs) in $B$ decays
are usually defined by $a_{CP}\propto \Gamma -\bar{\Gamma}$
and $A_{CP}(t)\propto \Gamma(t)-\bar{\Gamma}(t)$.
The former case, called direct CPA or CP-odd observable, needs
both weak CP violating and strong phases, while the latter one of the time
dependent CPA contains not only a non-zero CP-odd phase but also the
$B-\bar{B}$ mixing. We note that the present world average
for $a_{CP}^{\Psi K}$ is $0.79\pm0.12$  \cite{BCPV,SMBCPV}
comparing with the SM prediction of $0.70\pm0.10$ \cite{SMBCPV}.

To study CPV, one can also define some other useful observables by the
momentum correlations.
For example, in a three-body decay, the simplest ones are the
triple correlations of $\vec{s}\cdot ( \vec{p}_{i}\times \vec{p}_{j})$
\cite{Garisto-CQL}, where $\vec{s}$ is the spin carried by one of outgoing
particles and $\vec{p}_{i}$ and $\vec{p}_{j}$ denote any two independent
momentum vectors. Clearly, these triple momentum correlations are T-odd
observables since they change sign under the time reversal ($T$)
transformation of $t\rightarrow -t$. In terms of the CPT invariant theorem,
T violation (TV) implies CPV. Therefore, studying of T-odd observables can
also help us to understand the origin of CPV. We note that the T violating
observables such as the above triple correlations do not require strong
phases. In the decays of $B\rightarrow K^{*}l^{+}l^{-}$ ($l=e,\mu$, and $%
\tau $), the spin $s$ can be the polarized lepton, $s_{l},$ or the $K^{*} $
meson, $\epsilon ^{*}(\lambda )$. Considering the polarized lepton, as
known, the polarization is always associated with the lepton mass, and thus
we expect that the T violating effects are suppressed and less than $1\%$
for the light lepton modes \cite{geng-BRD}. Although the $\tau$ mode can
escape from the suppression, the corresponding branching ratio (BR) which is $%
O(10^{-7})$ is about one order smaller than those of $e$ and $\mu$ modes.

It is known that CPAs such as $a_{CP}$ in $B\to K^*l^+l^-$ are small
even with  weak phases of $O(1)$ due to the smallness of strong phases
\cite{KL}.
In this paper, we concentrate on the possibility of having large T-odd terms
such as $\vec{\epsilon}_{K^{*}}(\lambda )\cdot \left( \vec{p}_{l^{+}}\times
\vec{p}_{K^{*}}\right) $ $\propto \varepsilon _{\mu \nu \alpha \beta }q^{\mu
}\epsilon ^{*\nu }(\lambda )p_{l}^{\alpha }P^{\beta }$ in the decays of $%
B\to K^{*}l^{+}l^{-}$ $(l=e,\mu)$, and for simplicity we set $m_{l}=0.$ We
will show that these types of T violation are zero in the SM but they could
be sizable in new physics such as the theories with SUSY.

The effective Hamiltonian of $b\to sl^+l^-$ is given by~\cite{Buras}
\begin{eqnarray}
{\cal H} &=&\frac{G_{F}\alpha V_{tb}V_{ts}^{*}}{\sqrt{2}\pi }\left[ H_{1\mu
}L^{\mu }+H_{2\mu }L^{5\mu }\right] \,,  \nonumber \\
H_{1\mu } &=&C_{9}(\mu )\bar{s}\gamma _{\mu }(\mu )P_{L}b\ -\frac{2m_{b}}{%
q^{2}}C_{7}(\mu )\bar{s}i\sigma _{\mu \nu }q^{\nu }P_{R}b\,,  \nonumber \\
H_{2\mu } &=&C_{10}\bar{s}\gamma _{\mu }P_{L}b\,,\ L^{\mu }\,=\,\bar{l}%
\gamma ^{\mu }l\,,\ L^{5\mu }\,=\,\bar{l}\gamma ^{\mu }\gamma _{5}l\,
\label{heff}
\end{eqnarray}
where $V_{tq}\ (q=s,b)$ are the CKM matrix elements, $C_{i}\
(i=7,9,10)$ are the Wilson coefficients (WCs) and their
expressions can be found in Ref. \cite{Buras} for the SM. Since
the operator associated with $C_{10}$ is not renormalized under
the QCD, it is the only one with the $\mu $ scale free. Besides
the short-distance (SD) contributions, the main effect on the BR
comes from c\={c} resonant states such as $\Psi ,\Psi ^{\prime
}\,\,etc.,$ $i.e.$, the long-distance (LD) contributions. In
the literature \cite{DTP},
it has been suggested to
combine the factorization assumption (FA) and vector meson
dominance (VMD) approximation in estimating LD effects for the $B$
decays. Hence, we may include the
resonant effect (RE) by absorbing it to the related WC. The effective WC of $%
C_{9}$ is given by
\begin{eqnarray}
C_{9}^{eff} &=&C_{9}(\mu )+(3C_{1}(\mu )+C_{2}(\mu ))(h(x,s)+  \nonumber \\
&&\left. \frac{3}{\alpha^2}\sum_{j=\Psi ,\Psi ^{\prime }}k_{j}\frac{\pi
\Gamma (j\rightarrow l^{+}l^{-})M_{j}}{q^{2}-M_{j}^{2}+iM_{j}\Gamma _{j}}%
\right) \,,  \label{effc9}
\end{eqnarray}
where $h(x,s)$ describes the one-loop matrix elements of operators $O_{1}=%
\bar{s}_{\alpha }\gamma ^{\mu }P_{L}b_{\beta }\ \bar{c}_{\beta }\gamma _{\mu
}P_{L}c_{\alpha }$ and $O_{2}=\bar{s}\gamma ^{\mu }P_{L}b\ \bar{c}\gamma
_{\mu }P_{L}c$ \cite{Buras}, $M_{j}$ ($\Gamma _{j}$) are the masses (widths)
of intermediate states, and the factors $k_{j}$ $\sim -1/(3C_{1}(\mu
)+C_{2}(\mu ))$ are phenomenological parameters for compensating the
approximations of FA and VMD and reproducing the correct BRs $%
Br\left( B\rightarrow J/\Psi X\rightarrow l^{+}l^{-}X\right) =Br\left(
B\rightarrow J/\Psi X\right) $ $\times $ $Br\left( J/\Psi \rightarrow
l^{+}l^{-}\right) $. Here, we have neglected the small Wilson coefficients.
The transition amplitude for $B\rightarrow K^{*}l^{+}l^{-}$ decays is found
to be
\begin{eqnarray}
{\cal M}_{K^{*}}^{(\lambda )} &=&\frac{G_{F}\alpha _{em}V_{tb}V_{ts}^{*}}{2%
\sqrt{2}\pi }\left\{ {\cal M}_{1\mu }^{(\lambda )}L^{\mu }+{\cal M}_{2\mu
}^{(\lambda )}L^{5\mu }\right\}  \nonumber \\
{\cal M}_{a\mu }^{(\lambda )} &=&if_{1}\varepsilon _{\mu \nu \alpha \beta
}\epsilon ^{*\nu }(\lambda )P^{\alpha }q^{\beta }+f_{2}\epsilon _{\mu
}^{*}(\lambda )+f_{3}\epsilon ^{*}\cdot qP_{\mu }  \label{ampk*}
\end{eqnarray}
where subscript $a=1(2)$ while $f_{i}=h_{i}$ $(g_{i})$ ($i=1$, $2$, $3$) and
\begin{eqnarray}
h_{1} &=&C_{9}^{eff}(\mu )V(q^{2})-\frac{2m_{b}}{q^{2}}C_{7}(\mu )T(q^{2}),
\nonumber \\
h_{2(3)} &=&-C_{9}^{eff}(\mu )A_{0(1)}(q^{2})+\frac{2m_{b}}{q^{2}}C_{7}(\mu
)T_{0}(q^{2}),  \nonumber \\
g_{1} &=&C_{10}V(q^{2})\,,\ \ g_{2(3)}=-C_{10}A_{0(1)}(q^{2})\,.
\label{ampk1}
\end{eqnarray}
Here, $P=p_{1}+p_{2}$, $q=p_{1}-p_{2}$, and the definitions of the form
factors in Eq. (\ref{ampk1}) and the correspondences between our notations
and those used in the literature can be found in the Appendix of Ref. \cite
{CQ}.

To obtain the T-odd terms of $\varepsilon_{\mu\nu\alpha\beta}
q^{\mu}\epsilon^{*\nu}(\lambda)p_l^{\alpha}P^{\beta}$, we have to study the
processes of $B\rightarrow K^{*}l^{+}l^{-}\rightarrow (K\pi )l^{+}l^{-}$ so
that the polarization $\lambda $ and $\lambda ^{\prime }$ in the
differential decay rate, written as $d\Gamma \propto H(\lambda ,\lambda
^{\prime })$ ${\cal M}_{K^{*}}^{(\lambda )}$ ${\cal M}_{K^{*}}^{(\lambda
^{\prime })\dagger }$ with $H(\lambda ,\lambda ^{\prime })\equiv \epsilon
(\lambda )\cdot p_{K}$ $\epsilon ^{*}(\lambda ^{\prime })\cdot p_{K}$, can
be different. From Eq. (\ref{ampk*}), we see that ${\cal M}_{2\mu
}^{(\lambda )} $ only depends on $C_{10}$. Clearly, the T violating effects
can not be generated from ${\cal M}_{2\mu }^{(\lambda )} {\cal M}_{2\mu
^{\prime }}^{(\lambda ^{\prime })\dagger}$, but induced from ${\cal M}_{1\mu
}^{(\lambda )} {\cal M}_{1\mu^{\prime }}^{(\lambda ^{\prime })\dagger }$ and
${\cal M}_{1\mu}^{(\lambda )} {\cal M}_{2\mu ^{\prime }}^{(\lambda ^{\prime
})\dagger }$. This can be understood as follows: firstly, for the ${\cal M}%
_{1\mu }^{(\lambda )}{\cal M}_{1\mu ^{\prime }}^{(\lambda ^{\prime })\dagger
}TrL^{\mu }L^{\mu ^{\prime }}$ contributions with $TrL^{\mu }L^{\mu ^{\prime
}}\sim \left( p_{l^{-}}^{\mu }p_{l^{+}}^{\mu ^{\prime }}+p_{l^{-}}^{\mu
^{\prime }}p_{l^{+}}^{\mu }-g^{\mu \mu ^{\prime }}p_{l^{-}}\cdot
p_{l^{+}}\right) $, the relevant T-odd terms can be roughly expressed by
\begin{eqnarray}
&&{\cal M}_{1\mu }^{(\lambda )}{\cal M}_{1\mu ^{\prime }}^{(\lambda
^{\prime})\dagger }TrL^{\mu }L^{\mu ^{\prime }} \propto  \nonumber \\
&&Z_{1}{Im} h_{1}h_{3}^{*}\epsilon (0)\cdot q\varepsilon _{\mu \nu \alpha
\beta }q^{\mu }\epsilon ^{*\nu }(\pm )p_{l^{+}}^{\alpha }P^{\beta }+
\nonumber \\
&&Z_{2}{Im}h_{1}h_{2}^{*}\epsilon (0)\cdot p_{l^{+}}\varepsilon _{\mu \nu
\alpha \beta }q^{\mu }\epsilon ^{*\nu }(\pm )p_{l^{+}}^{\alpha }P^{\beta } +
\nonumber \\
&&Z_{3}{Im}h_{1}h_{2}^{*}\epsilon (\mp )\cdot p_{l^{+}}\varepsilon _{\mu \nu
\alpha \beta }q^{\mu }\epsilon ^{*\nu }(\pm )p_{l^{+}}^{\alpha }P^{\beta }
\label{im1}
\end{eqnarray}
where $Z_i\ (i=1,2,3)$ are constants. From Eq. (\ref{ampk1}), one gets ${Im}%
h_{1}h_{2}^{*}\sim{Im} h_{1}h_{3}^{*}\sim {Im}C_{9}^{eff} (\mu )C_{7}(\mu )$%
. We note that as shown in Eq. (\ref{im1}), the T-odd observables
could be non-zero if the process involves strong phases or
absorptive parts even without CP violating phases. By means of Eq.
(\ref{effc9}), $C_{9}^{eff}(\mu )$ includes the absorptive parts
such that the results of Eq. (\ref{im1}) are not vanished in the
SM. Secondly, for ${\cal M}_{1\mu}^{(\lambda )} {\cal M}_{2\mu
^{\prime }}^{(\lambda ^{\prime })\dagger }TrL^{\mu }L^{5\mu
^{\prime }}$, one gets
\begin{eqnarray}
&&\left( {\cal M}_{1\mu }^{(\lambda )}{\cal M}_{2\mu ^{\prime }}^{(\lambda
^{\prime })\dagger }+{\cal M}_{2\mu }^{(\lambda )}{\cal M}_{1\mu ^{\prime
}}^{(\lambda ^{\prime })\dagger }\right) TrL^{\mu }L^{5\mu ^{\prime }}
\propto  \nonumber \\
&& \left( {Im}h_{2}g_{3}^{*}-{Im}h_{3}g_{2}^{*}\right) \varepsilon _{\mu \nu
\alpha \beta }q^{\mu }\epsilon ^{*\nu }(\pm )p_{l^{+}}^{\alpha }P^{\beta }
\label{im2}
\end{eqnarray}
where $TrL^{\mu }L^{5\mu ^{\prime }}=-4i\varepsilon ^{\mu \mu ^{\prime
}\alpha \beta }q_{\alpha }p_{l^{+}\beta }$ has been used. From Eq. (\ref
{ampk1}), we find that ${Im}h_{2}g_{3}^{*}-{Im}h_{3}g_{2}^{*}$ is only
related to ${Im}C_{7}(\mu )C_{10}^{*}$ and the dependence of ${Im}C_{9}(\mu
)C_{10}^{*}$ is canceled in Eq. (\ref{im2}). For the decays of $%
b\rightarrow sl^{+}l^{-}$, since usually there are no absorptive parts in $%
C_{7}(\mu)$ and $C_{10}$, a non-vanishing value of ${Im}C_{7}(\mu
)C_{10}^{*} $ indicates the pure weak CP violating effects.

In order to derive the whole differential decay rate for the $K^{*}$
polarization, we choose $\epsilon (0)=(|\vec{p}%
_{K^{*}}|,0,0,E_{K^{*}})/m_{K^{*}}$,  $\epsilon (\pm )=(0,1,\pm
i,0)/\sqrt{2}$, and $%
p_{l^{+}}=\sqrt{q^{2}}(1,\sin \theta _{l},0,\cos \theta _{l})/2$ with $%
E_{K^{*}}=(m_{B}^{2}-m_{K^{*}}^{2}-q^{2})/2\sqrt{q^{2}}$ and $|\vec{p}%
_{K^{*}}|=\sqrt{E_{K^{*}}^{2}-m_{K^{*}}^{2}}$ in the $q^{2}$ rest frame and
$p_{K}=(1,\sin \theta _{K}\cos \phi ,\sin \theta _{K}\sin \phi ,\cos \theta
_{K})m_{K^{*}}/2$ in the $K^{*}$ rest frame where $\phi $ denotes the
relative angle of decaying plane between $K\pi $ and $l^{+}l^{-}$. The
differential decay rate with the only relevant terms is given by
\begin{eqnarray}
&&\frac{d\Gamma }{d\cos \theta _{K}d\cos \theta _{l}d\phi dq^{2}}=\frac{%
3\alpha _{em}^{2}G_{F}^{2}\left| \lambda _{t}\right| ^{2}\left| \vec{p}%
\right| }{2^{14}\pi ^{6}m_{B}^{2}}Br(K^{*}\rightarrow K\pi )  \nonumber \\
&&\times \{4\cos ^{2}\theta _{K}\sin ^{2}\theta _{l}\sum_{i=1,2}|{\cal M}%
_{i}^{0}|^{2}+\sin ^{2}\theta _{K}(1+\cos ^{2}\theta _{l})  \nonumber \\
&&\sum_{i=1,2}\left( |{\cal M}_{i}^{+}|^{2}+|{\cal M}_{i}^{-}|^{2}\right)
-\sin 2\theta _{K}\sin 2\theta _{l}\sin \phi  \nonumber \\
&&\sum_{i=1,2}{Im}\left( {\cal M}_{i}^{+}-{\cal M}_{i}^{-}\right) {\cal M}%
_{i}^{0*}-2\sin ^{2}\theta _{K}\sin ^{2}\theta _{l}\sin 2\phi  \nonumber \\
&&\sum_{i=1,2}{Im}\left( {\cal M}_{i}^{+}{\cal M}_{i}^{-*}\right) +2\sin
2\theta _{K}\sin \theta _{l}\sin \phi (Im{\cal M}_{1}^{0}  \nonumber \\
&&({\cal M}_{2}^{+*}+{\cal M}_{2}^{-*})-Im({\cal M}_{1}^{+}+{\cal M}_{1}^{-})%
{\cal M}_{2}^{0*})+\cdots ]\}\,,
\end{eqnarray}
where $|\vec{p}%
|=[((m_{B}^{2}+m_{K^{*}}^{2}-q^{2})/(2m_{b}))^{2}-m_{K^{*}}^{2}]^{1/2}$ and $%
{\cal M}_{i}^{0}$ and ${\cal M}_{i}^{\pm }$ denote the longitudinal and
transverse polarizations of $K^{*}$ respectively and their explicit
expressions are as follows:
\begin{eqnarray*}
{\cal M}_{a}^{0} &=&\sqrt{q^{2}}\left( \frac{E_{K^{*}}}{m_{K^{*}}}f_{2}+2%
\sqrt{q^{2}}\frac{\left| \vec{p}_{K^{*}}\right| ^{2}}{m_{K^{*}}}f_{3}\right)
, \\
{\cal M}_{a}^{\pm } &=&\sqrt{q^{2}}\left( \pm 2\left| \vec{p}_{K^{*}}\right|
\sqrt{q^{2}}f_{1}+f_{2}\right) ,
\end{eqnarray*}
The detailed derivation will be discussed elsewhere \cite{CQ2}.
Other distributions for the $K^{*}$ polarization can be found in
Ref. \cite{Kim}. From Eqs. (\ref{im1}) and (\ref{im2}),
we know that ${Im}({\cal M}_i^+-{\cal M}_i^-){\cal M}_i^{0*}$ and
${Im}({\cal M}_{i}^{+}{\cal M}_{i}^{-*})$ are from
${\cal M}_{1\mu }^{(\lambda )}{\cal M}_{1\mu ^{\prime
}}^{(\lambda ^{\prime })\dagger }TrL^{\mu }L^{\mu ^{\prime }}$ while
${Im}{\cal M}_{1}^{0}({\cal M}_{2}^{+*}+{\cal M}_{2}^{-*})-{Im}({\cal
M}_{1}^{+}+{\cal M}_{1}^{-}){\cal M}_{2}^{0*}$ is induced by ${\cal M}_{1\mu
}^{(\lambda )}{\cal M}_{2\mu ^{\prime }}^{(\lambda ^{\prime })\dagger
}TrL^{\mu }L^{5\mu ^{\prime }}$.

As seen from Eqs. (\ref{im1}) and (\ref{im2}), there are two possible
sources for T violation, which are related to ${Im}C_{9}^{eff}C_{7}^{*}$ and
${Im}C_{7}C_{10}^{*}$, respectively. In this paper, we only concentrate on
the contribution from ${Im}C_{7}C_{10}^{*}$ and explore the possibility of
existing new CP violating phases. To do this, we examine the T-odd
observable, defined by
\begin{equation}
\left\langle {\cal O}\right\rangle =\int {\cal O}d\Gamma \,.  \label{ov}
\end{equation}
where ${\cal O}$ is a T-odd five-momentum correlation, given by
\begin{equation}
{\cal O}=\frac{\left( \vec{p}_{B}\cdot \vec{p}_{K}\right) \left( \vec{p}%
_{B}\cdot (\vec{p}_{K}\times \vec{p}_{l^{+}})\right) }{\left| \vec{p}%
_{B}\right| ^{2}\left| \vec{p}_{K}\right| ^{2}\omega_{l^{+}} }%
\,
\end{equation}
with $\omega_{l^{+}}=q\cdot p_{l^{+}}/\sqrt{q^2}$. In the $K^{*}$
rest frame, we note that ${\cal O}=\cos \theta _{K}\sin \theta
_{K}\sin \theta _{l}\sin \phi $. The statistical significance of
the observable in Eq. (\ref{ov}) can be determined by
\begin{equation}
\varepsilon (q^{2})={\frac{\int {\cal O}d\Gamma }{\sqrt{(\int d\Gamma )(\int
{\cal O}^{2}d\Gamma )}}}\,.  \label{ss}
\end{equation}
Integrating all relevant angles in Eq. (\ref{ss}), we get
\begin{eqnarray}
\varepsilon (q^{2})&\simeq& \frac{0.76}{\sqrt{{\cal D}_{1}{\cal D}_{2}}} [Im%
{\cal M}_1^0({\cal M}_2^{+*}+{\cal M}_2^{-*})-  \nonumber \\
&& Im ( {\cal M}_1^++{\cal M}_1^-) {\cal M}_2^{0*}]\,,  \nonumber \\
{\cal D}_{1} &=&\sum_{i=1,2}\left[ \left| {\cal M}_{i}^{0}\right|
^{2}+\left| {\cal M}_{i}^{+}\right| ^{2}+\left| {\cal M}_{i}^{-}\right|
^{2}\right] \,,  \nonumber \\
{\cal D}_{2} &=&\sum_{i=1,2}\left[ \left| {\cal M}_{i}^{0}\right| ^{2}+\frac{%
1}{2}\left( \left| {\cal M}_{i}^{+}\right| ^{2}+\left| {\cal M}%
_{i}^{-}\right| ^{2}\right) \right] .
\end{eqnarray}
To observe the effect at the $n\sigma $ level, the required number of B
mesons is $N_{B}=n^{2}/(Br\cdot \varepsilon^2)$.

It is known that to study the exclusive decays of $B\to K^{*}l^{+}l^{-}$,
the main uncertainty is from the transition form factors in Eq. (\ref{ampk1}%
). The calculations of matrix elements for exclusive hadron decays
can be performed in the perturbative QCD (PQCD) approach developed
by Lepage-Brodsky (LB) \cite{BL}. However, with the LB approach,
it has been pointed out that perturbative evaluation of the pion
form factor suffers singularities from the end-point region with a
momentum fraction $x\to 0$ \cite{IL}.
 In order to take
care of the end-point singularities, the strategy of introducing
$k_{T}$, the transverse momentum of the valence quark, and
threshold resummations has been proposed and shown that the
end-point singularities can be dealt with self-consistent in the
PQCD \cite{MPQCD}.
Our calculations will base on such an approach. We use the results
that have been displayed in Ref. \cite{CQ}.
The form factors given by the other QCD approaches such as the
quark model (QM) and light-cone QCD sum rule (LCSR) can be found
in Ref. \cite{Bdecays,Aliev-MS}

It is well known that
supersymmetric theories not only supply an elegant
mechanism for the breaking of the electroweak symmetry
and a solution to the hierarchy problem, but provide
many new weak CP violating phases. Unfortunately, it has been
shown that with the universal soft breaking parameters, these
 phases are severely bounded  by electric dipole moments
(EDMs) of electron and neutron \cite{Garistosusy} so that the
contributions to $\epsilon$ and $\epsilon'$ become negligible.
However, one may avoid the EDM constraints by setting the
squark masses of the first two generations to be as heavy as few
TeV \cite{BKMW}. The SUSY models
with the non-universal soft A terms inspired by string theories
\cite{BK} and left-right symmetry \cite{BDM},
which have
unsuppressed
weak CP phases and lead to the observed $\epsilon$, $\epsilon'$
and the large CPA of $B\rightarrow X_{s}\gamma$, have also been
proposed.

To illustrate the new physics contributions, we shall use (a) a
model-independent approach by assuming that the CPV only arises
from $C_7$, the main effect for $b\rightarrow s\gamma$, with
taking $ImC_7=0.25$ and the values of remaining WCs are the same
as those in the  SM; and (b) the results of the generic
supersymmetric extension of the SM in Ref.
 \cite {Masiero} where,
instead of scanning the whole allowed parameter space, we take the
values
\begin{figure}[tbp]
\centerline{ \psfig{figure=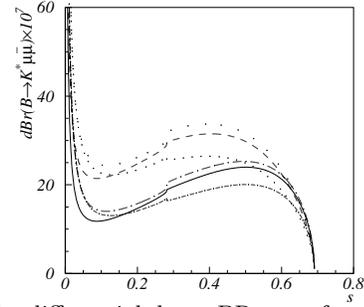, width=4.65cm} }
\caption{The differential decay BRs as a function of $s$: The
solid (dashed), long dash-dotted (dotted) and dash-dotted
(dense-dotted) lines describe the results of the PQCD (LCSR) in
the standard, model-independent with $ImC_7=0.25$ and SUSY models,
respectively.} \label{difrate}
\end{figure}
\begin{equation}
\begin{tabular}{lllllll}
$Re\,C_{7}^{SUSY}$ & $\simeq$ & $0,$& & $Im\,C_{7}^{SUSY}$ & $\simeq $ & $%
-0.27,$ \\
$Re\,C_{9}^{SUSY}$ & $\simeq $ & $-0.6,$ &  & $ImC_{9}^{SUSY}$ &
$\simeq $ &
$-0.1,$ \\
$Re\,C_{10}^{SUSY}$ & $\simeq $ & $8.5,$ &  & $Im\,C_{10}^{SUSY}$
& $\simeq $ & $-\;2.5\,.$
\end{tabular}
\label{npv}
\end{equation}
Here, the scale for WCs without specified is $M_W$, which can be taken
approximately the same as the SUSY scale.
Several remarks concerning on the SUSY models are given as follows:
(i) the source of the CP violating phases is embedded in the
sfermion mass matrices with the mass-insertion method \cite {Masiero};
(ii) we assume that the flavor diagonal terms in the SUSY models are
either real or small so that there are no new constraints from the EDMs of
neutron and electron;
(iii) for $Im\ C_7^{SUSY}=\pm 0.27$,
we find that $Br(B\rightarrow X_{s}\gamma)$ increases $\sim10\%$
by comparing with that in the SM \cite{KN},  consistent with the data of
$(3.22\pm0.40)\times 10^{-4}$ \cite{Expbsg}, and
 the rate CP asymmetry in $b\to
s\gamma$ is about $\pm 4\%$, much larger than the SM prediction
of $\sim 1\%$ \cite{SMbsg} but still within the recent $95\%$ range
of $-0.30$ to $0.14$ implied by the CLEO measurement \cite{CLEOCPV};
and (iv) the SUSY effect on $B\to J/\Psi\,K$ with the values in Eq.
(\ref{npv}) is small, while those with other allowed parameters in SUSY
have been discussed in Ref. \cite{SMBCPV}.

\begin{figure}[tbp]
\centerline{ \psfig{figure=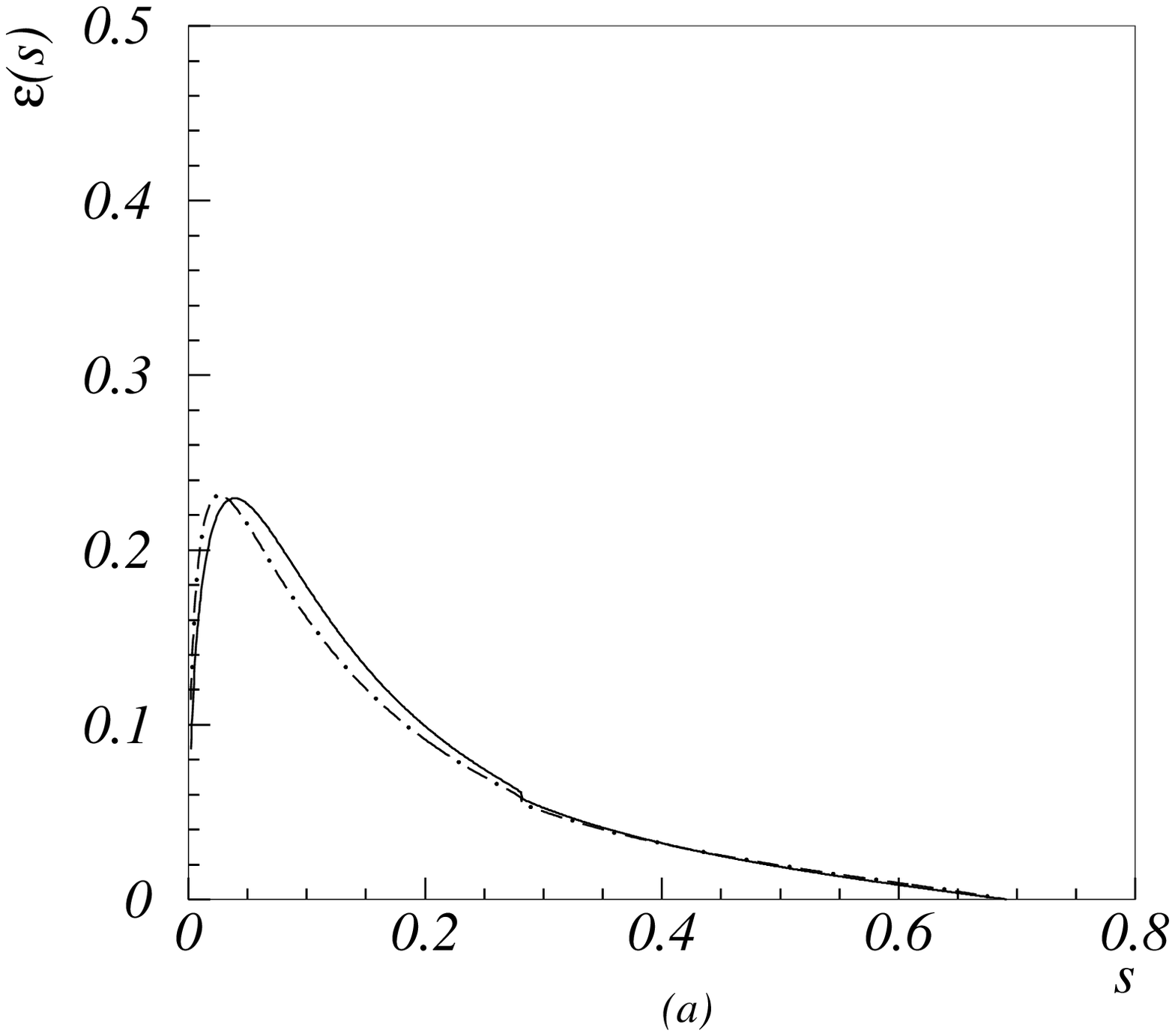, width=1.6in}
\psfig{figure=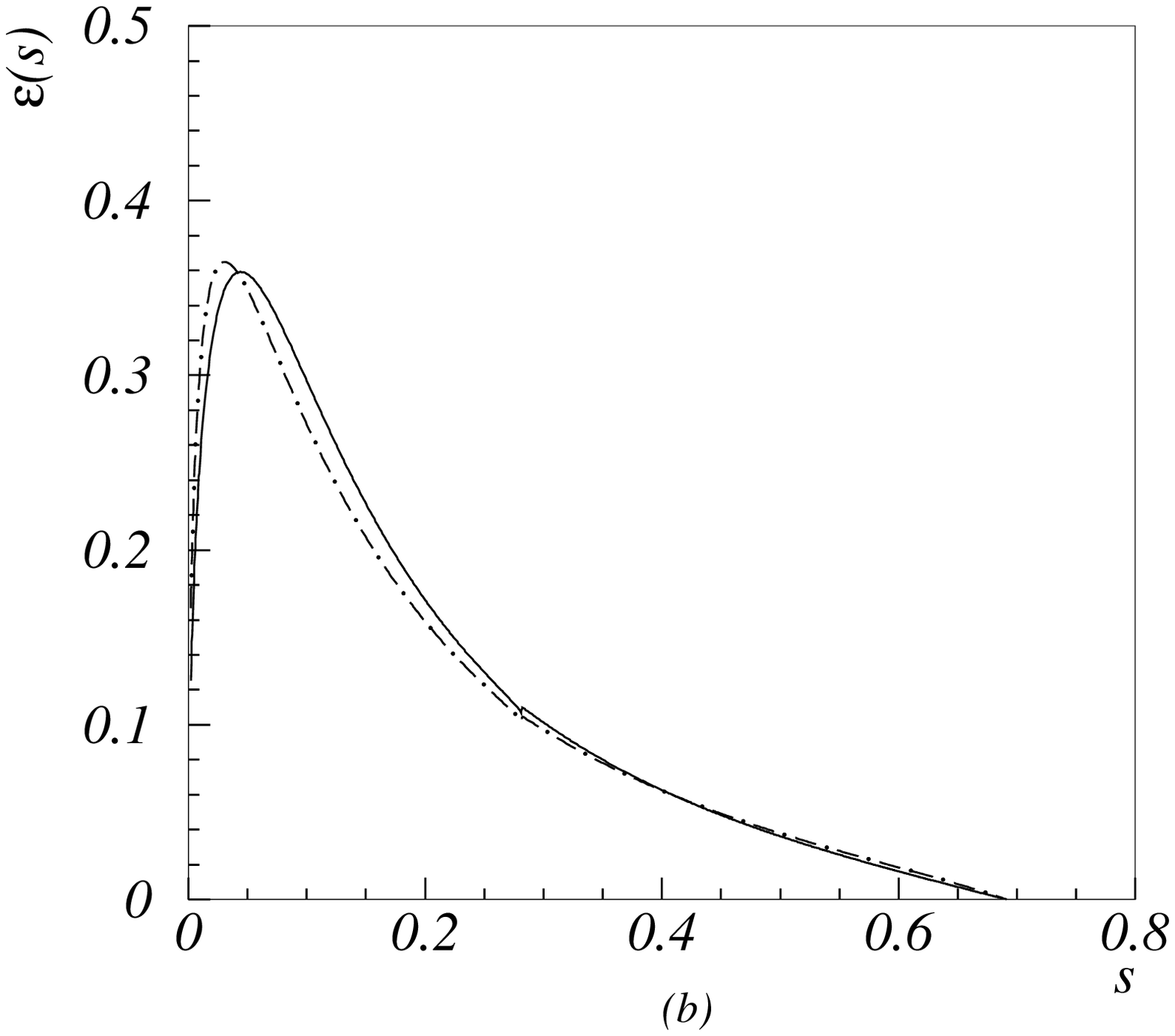, width=1.6in} } \caption{ The statistical
significance $\varepsilon$ of the T violating observable $\langle
{\cal O}\rangle$ in (a) the model-independent approach with
$ImC_7=0.25$ and (b) the generic SUSY model as a function of $s$.
The solid and dash-dotted lines stand for the results of the PQCD
and LCSR, respectively. } \label{tv}
\end{figure}

With the values
in Eq. (\ref{npv}), we find that the BRs of $B\to K^*l^+l^-$ in the
standard, model-independent and SUSY models
are $1.33\ (1.88)$, $1.51\ (2.06)$ and $1.29\ (5.17)\times 10^{-6}$
for the PQCD (LCSR), respectively.
We note that the corresponding BRs for $B\to Kl^+l^-$
are $5.32\ (5.17)$, $5.36\ (5.22)$ and $5.19\ (4.94)\times 10^{-7}$,
respectively, which so far are all consistent with the experimental data
\cite{Belle}.
The differential decay rates of $B\to K^*l^+l^-$ as a function of
$s=q^{2}/m^2_{B}$ are shown in Fig. \ref{difrate}, and the distribution of
$\varepsilon(q^2)$ is displayed
in Fig. \ref{tv}.
 From Fig. \ref{difrate}, we see that the uncertainty from the
different QCD approaches is compatible to the new physics effects.
Thus, it is not easy to ensure the existence of new physics by
only measuring the branching distribution. However, the T
violating observable $\langle {\cal O}\rangle$ which is zero in
the SM, could be large. In particular, in the SUSY model, the
statistical significance $\varepsilon (q^2)$ of $\langle {\cal
O}\rangle$ can be over $30\%$ at the low $q^{2}$ region and it is
insensitive to non-PQCD effects.

 Finally, we note that the origin of the T violating effect
 here would be quite different from the CPA in $b\to s\gamma$. For
example,
$\varepsilon (q^2)$ can be non-zero and large even for the case of
$Im\,C_7=0$
with CPV arising from phases in non-dipole WCs such as $C_{10}$
due to new physics,
whereas $a_{CP}(b\to s\gamma)$ is the same as that in the SM.

In summary, we have shown that, by measuring the angular distributions of $K$
and leptons, one can obtain the individual informations of longitudinal and
transverse polarizations of $K^{*}$ which are all sensitive to physics
beyond the SM. For reducing the uncertainties of QCD effects, one can define
some physical observables normalized by the differential decay rate. Among
different angular distributions, we have found that the T-odd contribution
arising from $M_{1\mu}^{(\lambda
)}M_{2\mu^{\prime}}^{(\lambda^{\prime})\dagger} TrL^{\mu}L^{5\mu ^{\prime }}$
is purely related to the weak CP violating phase, which could be sizable in
new physics such as the model with SUSY. Searching for such a T violating
distribution, one can distinguish the new CP violating source from the CKM
mechanism. We remark that to observe the T violating effects of $%
\langle {\cal O}\rangle$ in $B\to K^{*}l^{+}l^{-}\ (l=e,\mu)$ with $%
\varepsilon\sim 10\%$ at the $1\sigma$ level, at least $5\times 10^7\ B$
mesons are needed, which can be done at the B-factories in KEK and SLAC
and future hadron colliders. 

{\bf Acknowledgments:} This work was supported in part by the National
Science Council of the Republic of China under Contract Nos.
NSC-89-2112-M-007-054 and NSC-89-2112-M-006-033.

\end{document}